# Hybridization-brightened interlayer excitons in multilayer $WSe_2$ controlled by sliding ferroelectricity

Youngki Yeo[1,*], Idan Kizel[1,*], Dror Hershkovitz[1], Omri Meron[1], Iva Plutnarova[2], Zdenek Sofer[2], Watanabe Kenji[3], Takashi Taniguchi[4], Elad Katash[5], Dan Oron[5], Haim Suchowski[1], Jing-Yang You[6,†], Moshe Ben Shalom[1,†]

[1]School of Physics and Astronomy, Tel Aviv University, Tel Aviv, Israel

[2]Dept. of Inorganic Chemistry, University of Chemistry and Technology Prague, Technicka 5, 166 28 Prague 6, Czech Republic

[3]Research Center for Functional Materials, National Institute for Materials Science, Tsukuba, Japan

[4]International Center for Materials Nanoarchitectonics, National Institute for Materials Science, Tsukuba, Japan

[5]Department of Physics of Complex Systems, Weizmann Institute of Science, Rehovot 7610001, Israel

[6]Peng Huanwu Collaborative Center for Research and Education, Beihang University, Beijing 100191, China

[*]These authors contributed equally: Youngki Yeo, Idan Kizel

[†] Corresponding authors: Moshe Ben Shalom, Jing-Yang You

Interlayer excitons (IXs) combine spatially separated electron and hole wavefunctions with electrically tunable dipole moments, but their reduced wavefunction overlap generally limits optical emission, particularly for high-energy states. Here we report a bright, spectrally narrow high-energy interlayer exciton (h-IX) near 2.23 eV in few-layer $WSe_2$ and investigate how its optical activity evolves with layer number and sliding-ferroelectric polarization. In pristine 2H-$WSe_2$, we observe thickness-dependent brightening of high-energy exciton emission. GW–BSE calculations reveal mixed intralayer and interlayer configurations with enhanced optical weight, suggesting that multilayer hybridization contributes to the observed brightening. In parallel-stacked 3L/3L $WSe_2$ containing a sliding-ferroelectric interface, the h-IX exhibits polarization-dependent energies and Stark responses that distinguish $P_{up}$ and $P_{down}$ domains. Electric-field-driven domain-wall motion produces hysteretic transfer of spectral weight between the two h-IX branches, enabling optical tracking of ferroelectric switching. These results establish multilayer hybridization as a route to optically active high-energy interlayer excitons and sliding ferroelectricity as a complementary degree of freedom for controlling their energy, dipole and emission.

## 1. Introduction

Interlayer excitons (IXs), where electrons and holes are spatially separated in different layers, have emerged as a fascinating platform for exploring many-body physics and optoelectronic applications due to their unique properties distinct from intralayer excitons[1–4]. Oriented electron–hole pairs localized in different layers carry an out-of-plane dipole and couple to external electric fields. The static dipole of a layer-localized pair increases with electron-hole separation, while the reduced wavefunction overlap suppresses the optical transition matrix element and hence the oscillator strength[5,6]. Resonant photocurrent measurements place the IX oscillator strength roughly two orders of magnitude below the intralayer value[5], consistent with the order-of-magnitude longer IX photoluminescence lifetimes measured in the same material system[3] and with microscopic calculations of the interlayer transition matrix element[7]. Such control over interlayer excitonic states in vdW heterostructures is central to optoelectronic applications such as solar cells[8], and photocatalysis[9], exciton transistor[10] and fundamental studies such as Bose-Einstein condensation[11] and valleytronics[12].

Photoluminescence (PL) in transition metal dichalcogenides (TMDs) is typically dominated by the lowest-energy exciton, with much weaker emission at higher energies[13], reflecting efficient relaxation from higher energy states via ultrafast scattering processes[14–16]. Nevertheless, optically accessible high-lying interlayer excitons have recently been identified in bilayer $WSe_2$, where their spatially indirect character was established through pronounced Stark splitting and switchable out-of-plane dipoles[17]. In molecularly gated TMD homobilayers, a high-energy interlayer exciton becomes optically visib le when a strong electric field brings it into resonance with a bright intralayer exciton[18]. More generally, the optical properties of interlayer excitons are highly sensitive to their local electronic environment. In TMDs heterostructures, moiré potentials and interlayer coupling can reshape interlayer exciton emission by shifting exciton energies[19], splitting peaks[20], and producing new emission features[21]. In multilayer homostructures, however, additional layers introduce multiple intralayer and interlayer electron–hole configurations that can mutually hybridize. How this enlarged configuration space modifies the balance between optical activity and out-of-plane charge separation, and whether such multilayer hybridization can enhance high-energy interlayer-exciton emission, remains largely unexplored.

Recently, sliding ferroelectricity in two-dimensional materials has introduced a novel degree of freedom for manipulating these excitonic states. In contrast to conventional ferroelectrics, whose polarization arises from ionic displacement, sliding ferroelectricity originates from the broken inversion symmetry of parallel-stacked van der Waals layers, producing a spontaneous out-of-plane polarization that can be switched by the in-plane sliding of adjacent layers[22,23]. The resulting asymmetric interlayer coupling imprints distinct optical signatures on interlayer excitons, which have been used to resolve sliding ferroelectric states and their switching dynamics[24,25]. Beyond readout, however, the large built-in fields intrinsically generated by sliding ferroelectricity offer an opportunity to control high-energy excitonic states by modifying their energies and layer hybridization.

Here we report a bright high-energy interlayer exciton (h-IX) near 2.23 eV in few-layer $WSe_2$ and show that its optical activity and dipolar character can be tuned through layer thickness and sliding-ferroelectric polarization. In pristine 2H-$WSe_2$, the h-IX emission stays weak from 3L to 5L and becomes markedly stronger in 6L despite a reduction in its Stark-derived electron–hole separation. GW–BSE calculations show enhanced intrinsic optical weight and global electron hole correlations spanning multiple layers. In parallel-stacked 3L/3L $WSe_2$, the sliding-ferroelectric interface further splits the h-IX response into polarization-dependent branches whose energies, Stark shifts and intensities track $P_{up}/P_{down}$ domains and their electric-field-driven switching. This work demonstrates that structural symmetry engineering in TMD homostructure devices can enable the coexistence of strong optical activity and electrically controllable excitonic dipoles, establishing multilayer hybridization as a route to bright high-energy interlayer excitons in the visible range.

## Layer thickness dependence of the interlayer exciton

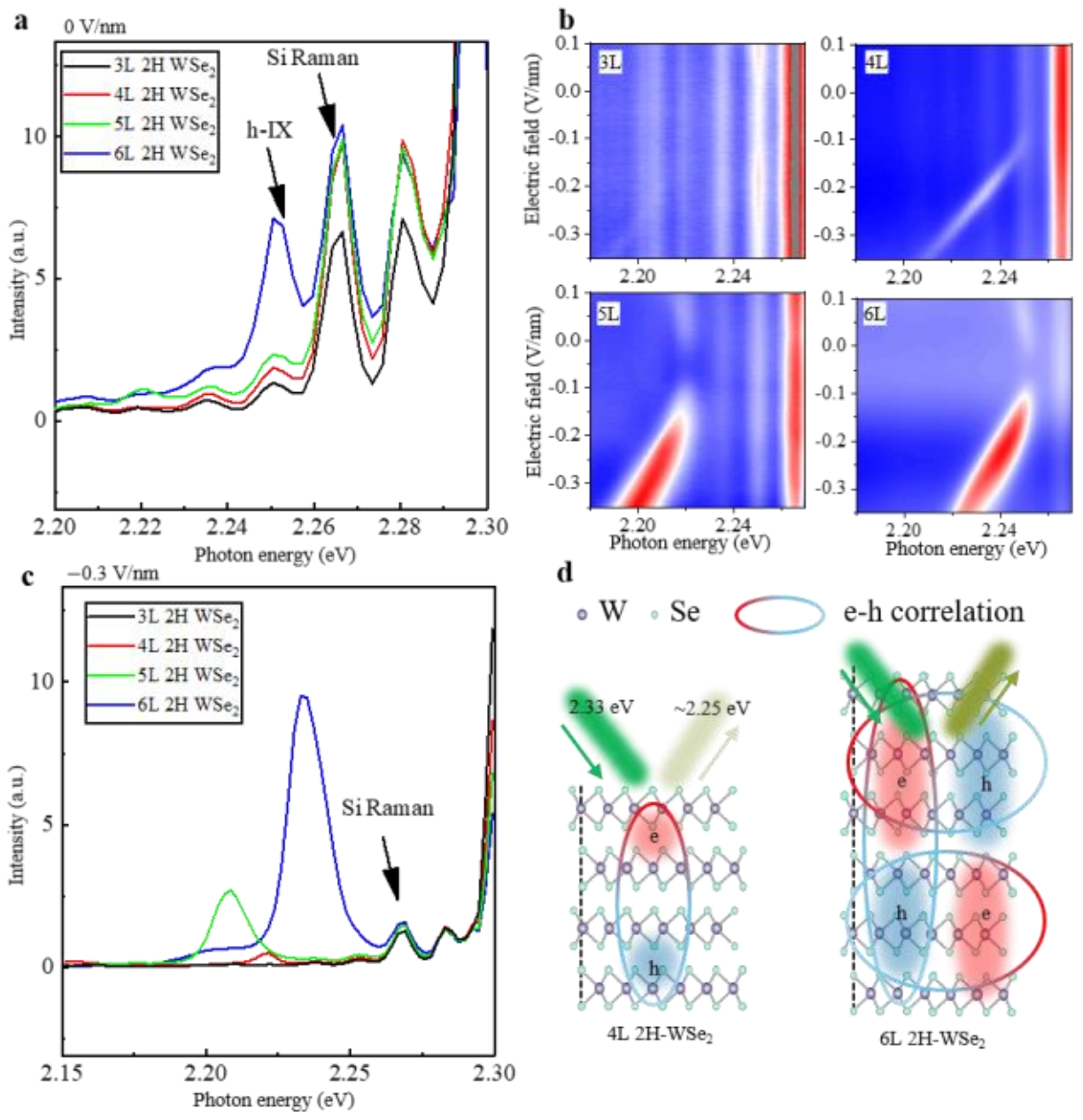

**Figure 1. Thickness-dependent brightening and Stark tuning of h-IX in multilayer 2H-$WSe_2$. a,** PL spectra of 3L–6L 2H-$WSe_2$ at $E_z$ = 0 V/nm. The h-IX peak near 2.25 eV grows markedly with layer number and is strongest in 6L, while the Si Raman line at ~2.27 eV remains nearly unchanged. **b**, PL intensity maps as a function of out-of-plane electric field $E_z$ and photon energy for 3L, 4L, 5L and 6L. **c**, PL spectra at $E_z$ = −0.3 V/nm. **d**, Schematic of the electron–hole (e–h) correlation in 4L and 6L 2H-$WSe_2$. Red and blue clouds denote electron and hole densities, and ovals mark correlated e–h pairs. In 4L, h-IX is dominated by layer-separated e–h configuratio ns. The schematic illustrates layer-separated and intralayer e-h configurations, with a more extended corr elation in 6L. The intralayer component provides optical strength. Green arrows indicate 2.33 eV excitation and ~2.25 eV h-IX emission, with shading reflecting relative emission intensity. All spectra were measured at $T$=4 K.

We measured the PL spectra and the Stark effect of pristine 3L–6L 2H-$WSe_2$ (Fig. 1). The device consists of h-BN/bilayer graphene/28 nm h-BN/3–6L $WSe_2$/10 nm graphene, and an optical image is shown in Supplementary Fig. S1. Among the 3L–6L flakes, the h-IX peak becomes most pronounced in 6L 2H-$WSe_2$ (Fig. 1a). The dipole character of the h-IX is particularly evident in its Stark response (Fig. 1b). A clear evolution in the h-IX peak intensity is observed as a function of layer number within the photon energy range of 2.18–2.27 eV (Fig. 1b), with the strongest emission in 6L. The sharp peak near 2.27 eV, arising from a Raman mode of the Si substrate and independent of $WSe_2$ thickness, serves as a reference for comparing these h-IX intensities across layer thicknesses. At −0.3 V/nm, the h-IX intensity increases with layer number, with the strongest emission in 6L (Fig. 1c). The more gradual layer dependence under applied field than at zero field may reflect the asymmetric environment of the thinner flakes, in which the vertical field displaces the h-IX away from the graphene-contacted bottom layer and thereby suppresses interfacial quenching.

The linear Stark shift allows us to extract the effective electron–hole separation $d$ through $|\Delta E| = ed\frac{\epsilon_{\mathrm{hBN}}}{\epsilon_{\mathrm{WSe_2}}}|E_z|$, where $\Delta E$ is the Stark shift, $E_z$ is the vertical electric field, d is the effective electron-hole separation, $e$ is elementary charge, and $\epsilon_{hBN} \approx 3$ and $\epsilon_{WSe_2} \approx 7$ are the out-of-plane dielectric constants of h-BN and $WSe_2$, respectively[26] . The factor $\epsilon_{hBN}/\epsilon_{WSe_2}$ accounts for the dielectric cor rection to the field inside $WSe_2$. In 3L, the h-IX is barely discernible even at the highest fiel ds, whereas a resolvable feature appears in 4L and becomes especially pronounced in 6L. In 4L, the Stark slope and net charge separation distance of the h-IX are estimated to be 0.165 e·nm and 3.85 Å. Interestingly, for the 5L and 6L 2H-$WSe_2$ devices, the Stark slope decreases to 0.0848 e·nm and 0.107 e·nm. Incorporating the same dielectric constant yields a corrected charge separation distance of 1.98 Å and 2.50 Å, which is notably smaller than that of the 4L device. The reduced effective dipole in 5L and 6L reflects a smaller separation between the electron and hole charge centers, which m ultilayer hybridization can produce through redistribution of the layer weights. The redistribution of charge centers supports the oscillator strength and PL brightening at 6L. The oscillator strength of an interlayer exciton is governed by the spatial overlap between the electron and hole wavefunctions, which is suppressed when the two states remain largely layer-localized[5,27–29]. Mixing with intralayer configurations can provide optical strength to an interlayer-dominated exciton through its intral ayer transition amplitude[6,28–30]. The linear fits used to extract the slope and its standard errors are shown in Supplementary Fig. S2.

In multilayer $WSe_2$, photoexcited carriers relax toward the indirect gap, so emission from higher-lying states is intrinsically weak and the direct-gap A exciton emission is progressively quenched with increasing thickness, and emission from higher-energy states is typically two to three orders of magnitude weaker than band-edge emission[31]. We attribute this opposite trend to intra-interlayer hybridization (Fig. 1d), which enhances the radiative strength of the h-IX in thicker flakes, as supported by the Stark measurements and GW-BSE calculations below.

Spectrally filtered time-resolved photoluminescence measurements (see methods section) reveal a more prominent delayed contribution in 4L than in 5L, 6L, and 3L/3L $WSe_2$ (Supplementary Section S3 and Fig. S3). This evolution may accompany changes in exciton population and relaxation pathways with thickness. Because the detection windows do not isolate the h-IX, these measurements do not determine its radiative lifetime.

## GW-BSE calculations for interlayer exciton brightening

To determine the microscopic origin of the thickness-dependent h- IX brightening, we performed spinor GW–BSE calculations for 4L and 6L 2H-$WSe_2$ (Supplementary S e c t i o n S 4). Both systems exhibit a bright excitonic manifold near 2. 3 4 eV (Fig. 2 a), composed of many closely spaced BSE eigenstates with finite transition dipoles. For a quantitative comparison, we multiplied the imaginary part of the BSE macroscopic dielectric function by the super cell height to obtain the sheet optical response. The peak-centered ±60 meV windows, 2.2771–2.397 1 eV for 4L and 2.2878–2.4078 eV for 6L, contain 544 and 1273 BSE eigenstates, respectively, c ounted without an optical-strength cutoff. The window-summed squared transition dipoles increa se by a factor of 1.46 from 4L to 6L.

Figure 2b resolves the momentum-space composition of the 6L bright manifold. The BSE pair weights were averaged over the same analysis window using the squared transition dipoles and normalized over the full Brillouin zone. Both K/K′ and Λ(Q) regions contribute to the exciton wavefunction s. The valley-resolved optical amplitudes show that Q-region self-terms provide 76.49% and 7 1.11% of the summed dipole strength in the 4L and 6L windows, respectively, while construc tive K–Q interference contributes a further 12.22% and 14.09% (Supplementary Table S2). Th e largest optical channel is the Q-region intralayer component, while the optically weighted K

-region interlayer probability increases from 12.99% in 4L to 30.41% in 6L. These vertical electron-hole transitions combine strong intralayer optical coupling with substantial interlayer character.

Figure 2c relates optical strength to the global interlayer probability, obtained by integrating the joint electron-hole density over both particle coordinates and all spinor components. Interlayer-dominated states, defined by $W_{inter} > W_{intra}$, carry 10.01% of the window-summed dipole strength in 4L and 31.49% in 6L. Their summed strength increases by a factor of 4.60, comparable in scale to the observed 4-5-fold PL enhancement. This redistribution of optical weight into interlayer-dominated states provides a microscopic basis for the enhanced h-IX emission in 6L. The strongest interlayer-dominated state in each window, labelled 4L h-IX and 6L h-IX, contains both intralayer and interlayer components (Supplementary Fig. S4).

The out-of-plane dipole depends on the signed electron–hole separation. Intralayer configurations contribute little to this separation, while oppositely oriented interlayer configurations can contribute with opposite signs. To examine the field response, we coupled the zero-field BSE exciton basis to an out-of-plane electric field through the electron–hole position-difference operator (Supplementary Section S4.6). Fits to selected energy-ordered model levels over the common internal-field interval −0.15 to −0.06 V nm$^{-1}$ give effective Stark lengths of 3.45 Å for 4L and 2.72–2.76 Å for 6L (Fig. 2d), compared with experimental values of 3.85 and 2.50 Å. The model captures the experimental Å-scale separation and the smaller 6L value over this interval (Supplementary Fig. S5).

The optical decomposition in Fig. 2e separates the window-summed dipole strength into intralayer and interlayer self terms and their coherent cross term. The intralayer term increases from 4745.21 in 4L to 6923.10 in 6L, the interlayer term from 12.24 to 20.24, and the cross term from 439.53 to 673.37, in common calculated dipole-strength units. The intralayer and cross terms account for 90.11% and 9.68% of the total increase, respectively (Supplementary Table S3). For the representative h-IX states (Fig. 2f), the total strength increases from 32.73 to 401.79, while the intralayer probability increases from 14.89% to 37.56%. The mixed excitons therefore acquire optical strength predominantly through their intralayer transition amplitudes, with a constructive contribution from intralayer–interlayer interference. This explains how inter layer-dominated states can become optically bright in 6L. The calculated strengths quantify op tical coupling, while the measured PL also depends on exciton populations and relaxation pat hways.

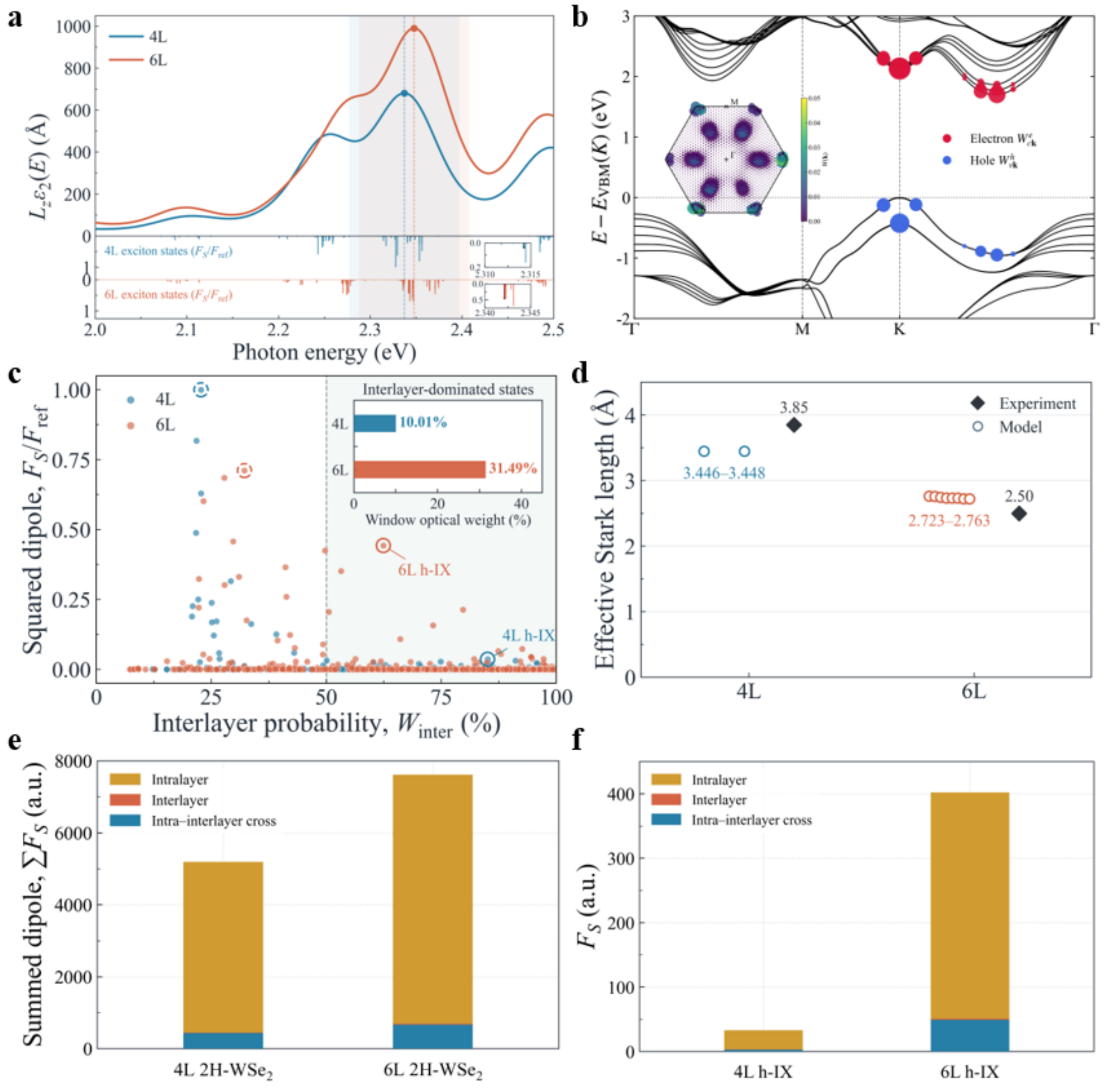

**Figure 2. Layer character, optical activity and Stark response of high-energy excitons in multilayer $WSe_2$.** **a**, Sheet optical response $L_z\varepsilon_2(E)$ for 4L and 6L with a common Voigt broadening of $\sigma = \gamma = 20$ meV. Here, $L_z$ is the supercell height, $\sigma$ is the Gaussian standard deviation and $\gamma$ is the Lorentzian half-width. Shading marks the peak-centred ±60 meV analysis windows. Downward sticks show squared transition dipoles $F_S/F_{ref}$, using the s trongest 4L state in its window as the common reference. Insets resolve the representative h-IX states. **b**, GW

quasiparticle bands of 6L with electron (red) and hole (blue) marginal weights averaged over its window with $F_S$ weighting. Energies are referenced to the valence-band maximum at K. The inset shows the normalized momentum distribution. **c**, Squared transition dipoles versus global interlayer probability for all 544 and 1,273 states in the 4L and 6L windows. The inset gives the optical share of interlayer-dominated states, defined by $W_{inter} > W_{intra}$. Solid-outline circles mark the strongest interlayer-dominated states; dashed circles mark the strongest optical transitions in each window. **d**, Effective Stark lengths from selected energy-ordered levels of the z ero-field BSE model over -0.15 to -0.06 V nm$^{-1}$ internal field (open circles), compared with dielectric-c orrected experiment (diamonds). Horizontal offsets separate the selected model levels. **e**,**f**, Absolute intralayer and interlayer optical self terms and their coherent cross terms summed over the full windows **e** a nd evaluated for representative 4L h-IX and 6L h-IX **f**.

## Polytype dependent interlayer excitons

Beyond thickness-driven hybridization, the h-IX can be further modulated by the built-in field of a polar 3R-type interface. Stacking two 3L 2H-$WSe_2$ flakes in parallel orientation creates a rhombohedral (3R-like) interface that breaks inversion symmetry (point group $C_{3v}$), driving interlayer charge transfer and a spontaneous out-of-plane polarization at the junction between the two 3L blocks[32–34]. The direction of this polarization is set by the local atomic registry: polarization-up (down) domains form when a W (Se) atom sits directly above a Se (W) atom[33] (Fig. 3a). We fabricated such a near-parallel 3L/3L $WSe_2$ homostructure, encapsulated by h-BN and ~10 nm graphene, on a 90 nm-thick $SiO_2$/Si substrate (Supplementary Fig. S1) to investigate how h-IX at 2.23 eV couples to this sliding ferroelectric polarization.

We observe a robust spatial correlation between the electrical and optical signatures of these domains. To confirm the high spatial resolution and the correlation between the h-IX and the sliding ferroelectric domains, we introduced bar- and circle-shaped holes a few hundred nanometers in size into bilayer graphene to confine the sliding ferroelectric domains (Supplementary Fig. S6). The $P_{up}$ domain, which shows a higher surface potential in the KPFM map, coincides with the bright regions in the integrated high-energy PL map (2.22–2.24 eV). The KPFM image is taken before putting top graphene electrodes and the high-energy PL map was taken after applying a field of 0.214 V/nm (Fig. 3b). We extracted the broad PL spectra (1.5 eV-2.7 eV) measured under 532 nm continuous-wave excitations at 4K to comparatively analyze the two domains where the extracted spots are marked as red and blue circles in Figure 3b (Fig. 3c, inset). We can mainly observe three peaks which are $KK_A$, up conversion photoluminescence (UPL) and the h-IX. $KK_A$ transition is the direct (K-K) band-edge transition of the A exciton which is dominant in few layers $WSe_2$. Because the A-exciton energy and intensity are sensitive to the local carrier density, polarization-induced electrostatic doping makes it a sensitive optical probe of the ferroelectric domains[35,36]. In practice, however, the A-exciton response also reflects strain[37], defects[38], and dielectric-environment variations[39,40], which complicates its use as a clean probe of ferroelectric polarization. Crucially, the A-exciton signal weakens with increasing layer thickness as relaxation into lower-energy momentum-indirect states reduces the population contributing to the K-K emission[41,42]. As a result, the A-exciton spectra provide little contrast between the upward and downward polarization states (Fig. 3c). The peak near 2.6 eV is attributed to upconversion photoluminescence (UPL) from a Γ–Γ transition, generated through exciton–exciton annihilation of Λ–Γ and Λ′–Γ dark excitons, which was recently observed to be polytype-dependent[43,44].

Finally, we observe bright, narrow emission lines emerging near 2.25 eV, well above the A exciton energy (Fig. 3c). The h-IX emission is polarization-selective: the $P_{up}$ and $P_{down}$ domains exhibit distinct h-IX branches (h-$IX^+$ and h-$IX^-$), which appear at 2.234 eV and 2.245 eV, respectively (Fig. 3c). Owing to the 11 meV splitting, the h-IX provides a more direct optical fingerprint of the underlying ferroelectric domains than the A-exciton. The h-IX transition energy responds to the difference between the electrostatic potentials averaged over its electron and hole distributions. Multilayer hybridization can place part of both distributions on common layers, allowing their electrostatic contributions to the transition-energy shift to cancel. Intralayer components contribute little to the out-of-plane dipole, while oppositely oriented interlayer components can further reduce the net response to the interfacial potential.

The layer-thickness-dependent h-IX intensities of 2H-$WSe_2$ are also reflected in marginally twisted 2L/2L $WSe_2$. The $P_{up}$ and $P_{down}$ domains are clearly distinguished in KPFM measurements (Supplementary Fig. S7), yet the h-IX emission extracted from both regions is substantially weaker than in the thicker 2H structures. The weaker h-IX emission in the 2L/2L structure may reflect less effective hybridization between layer-separated electron-hole configurations and optically bright intralayer excitations. This weaker mixing would limit the optical strength acquired by the h-IX through its intralayer components[5,27,29].

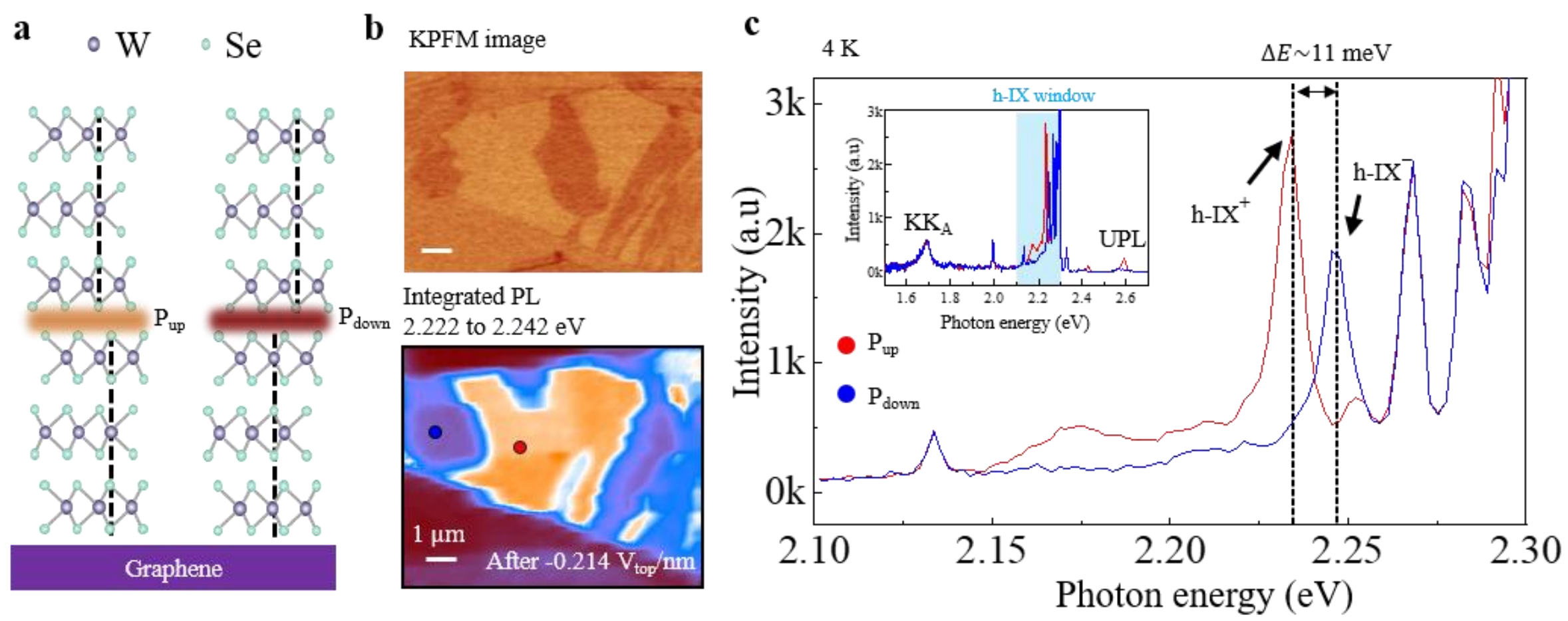

**Figure 3**. **Polytype-dependent optical response of high-energy interlayer excitons in 3L/3L $WSe_2$ homostructures. a,** Schematics of the polytypes with upward ($P_{up}$) and downward ($P_{down}$) spontaneous polarizations for $WSe_2$ on graphene. **b,** Spatial maps of the surface potential measured by Kelvin probe force microscopy (KPFM) (top) and the integrated photoluminescence (PL) intensity of the high-energy interlayer exciton peak (2.22–2.24 eV) (bottom). The PL image was taken after applying −4 V (−0.214 V/nm) to the top gate, showing expansion of the $P_{up}$ domain. **c,** Representative PL spectra (2.10–2.30 eV) extracted from the $P_{up}$ (red) and $P_{down}$ (blue) domains marked in panel **b**; h-$IX^+$ and h-$IX^-$ denote the $P_{up}$ and $P_{down}$ domains, respectively. The inset shows the full PL spectrum of 3L/3L $WSe_2$, comprising the $KK_A$, h-IX, and upconversion photoluminescence (UPL) features.

## Probing ferroelectric switching with high energy interlayer excitons

We investigate the switching dynamics of h-IXs in 3L/3L $WSe_2$ encapsulated by graphene and h-BN (Fig. 4). Sliding ferroelectric switching is mediated by pre-existing domain wall movement[22]. We initially positioned the excitation spot within a $P_{down}$ domain adjacent to a $P_{up}$ domain to capture PL spectrum with domain wall sliding. All measurements are performed at 4 K. As the negative electric field is swept, the h-$IX^-$ exhibits a pronounced redshift with a Stark slope of 0.209 e nm. After accounting for dielectric screening, this slope corresponds to an effective out-of-plane electron–hole separation of 4.88 Å. This value exceeds the effective dipole lengths of approximately 2-4 Å observed for h-IXs in reference 2H-$WSe_2$ multilayers, indicating an enhanced out-of-plane excitonic dipole in the sliding-ferroelectric structure. This enhancement may reflect the sliding-induced change in stacking registry, which breaks inversion symmetry and modifies the layer hybridization and vertical charge distribution of the h-IX. The large effective separation therefore provides an electronic signature of the strongly layer-polarized excitonic state associated with the polar stacking configuration.

Around -0.12 V/nm, a pronounced transfer of spectral weight from h-$IX^-$ to h-$IX^+$ is observed (Fig. 4a), consistent with the propagation of a domain wall through the excitation region and the expansion of the $P_{up}$ domain. As switching proceeds, the intensity of the h-$IX^-$ peak decreases while that of h-$IX^+$ increases. In contrast to h-$IX^-$, the h-$IX^+$ state exhibits a weaker Stark response, indicating a smaller effective out-of-plane excitonic dipole and/or stronger electrostatic screening in the polarization-up configuration. During the reverse field sweep, the inverted intensity contrast between h-$IX^+$ and h-$IX^-$ is retained (Fig. 4b), demonstrating the hysteretic nature of the polarization switching.

Upon sweeping the electric field in the positive direction, switching from the $P_{up}$ to the $P_{down}$ state occurs around 0.18 V/nm (Fig. 4c). The magnitude of the h-$IX^+$ Stark slope in this field direction is 0.141 e·nm, corresponding, after applying the same dielectric correction, to an effective electron–hole separation of 3.29 Å. This value is substantially smaller than the 4.88 Å obtained during the negative-field sweep, revealing an asymmetric excitonic response to the two field polarities. Such asymmetry may arise in part from field-dependent charge redistribution and screening near the graphene/$WSe_2$ interface. The resulting change in the local electrostatic environment can reduce the effective vertical dipole of the h-IX. On the subsequent return sweep, h-$IX^-$ remains the dominant state while the h-$IX^+$ emission remains strongly suppressed (Fig. 4d), further demonstrating the remanent character of the polarization-down configuration.

In asymmetric device geometry, band alignment with graphene, doping, and electric field could alter the band structure of $WSe_2$. To reduce these effects, we prepared a symmetric h-BN/graphene/h-BN/3L 2H-$WSe_2$/3L 2H-$WSe_2$/h-BN/graphene structure (Supplementary Fig. S1b). Its KPFM image shows a large ferroelectric domain (Supplementary Fig. S8a). At 0 V/nm, the polarization-up and -down domains are indistinguishable in PL. At 0.066 V/nm, we can clearly observe the two domains. The spectrally integrated h-IX PL map reveals stronger emission from the $P_{up}$ domain than from the $P_{down}$ domain. We stationed the beam at the polarization-down domain and swept the electric field from negative to positive values. The h-IX$^-$ peak changes little in the negative-field regime, and the slope reverses around 0.02 V/nm. Backward sweep reversed the polarization around 0.01 V/nm. The Stark slope of h-IX$^+$ in the symmetric device is −0.127 e·nm, corresponding, after accounting for dielectric screening, to an electron–hole separation of 2.97 Å. For h-IX$^-$, the slope is 0.202 e·nm, corresponding to a separation of 4.71 Å, comparable to that measured in the asymmetric device.

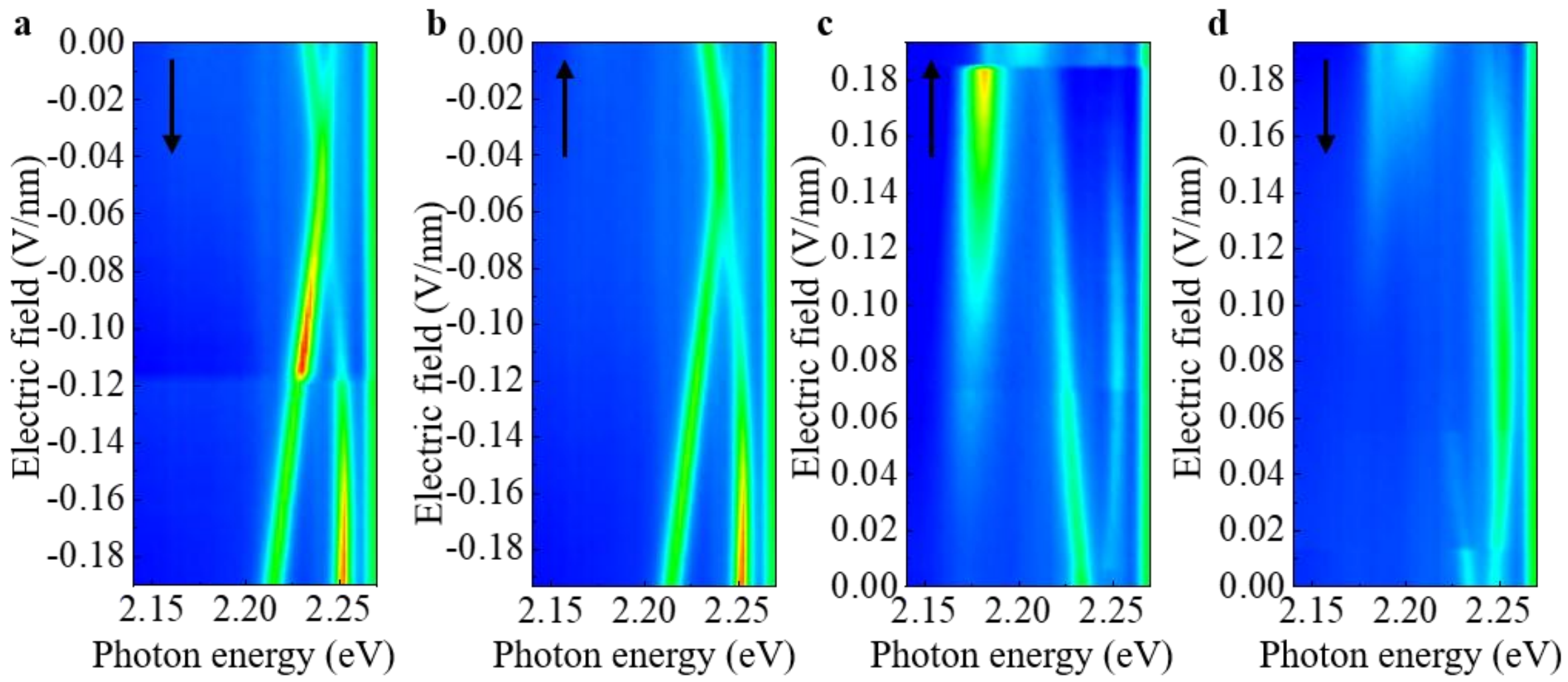

**Figure 4. Switching dynamics with high energy interlayer excitons. a-d**, Photoluminescence (PL) intensity maps as a function of photon energy and out-of-plane electric field, capturing a complete polarization switching loop. The black arrows indicate the chronological direction of the electric field sweep sequence. **a,** Spectral evolution during the downward sweep into the negative field regime, tracking the initial Stark shift trajectory. **b,** Return sweep from negative saturation back to zero field. **c,** Upward sweep into the positive field regime, highlighting the threshold field for deterministic domain switching and subsequent exciton state modification. **d,** Downward return sweep from positive saturation back to zero field, completing the closed hysteresis loop.

## Conclusion

The non-monotonic thickness dependence of the h-IX dipole moment reveals that exciton brightness and electron–hole charge separation evolve differently with increasing $WSe_2$ thickness. While the h-IX emission is strongest in 6L, the effective separation reaches approximately 3.85 Å in 4L and decreases to about 1.98–2.50 Å in 5L and 6L. Our GW–BSE calculations reveal coexisting intralayer and interlayer contributions to the bright excitonic states. The calculated window-summed optical streng th of interlayer-dominated states increases by a factor of 4.60 from 4L to 6L. Their optical a mplitudes are predominantly carried by Q-region intralayer components, while the frozen-basis Stark model illustrates a reduced effective dipolar response over the selected field interval. Th ese results support the experimentally observed combination of stronger h-IX emission and sm aller Stark-derived separation.

Sliding ferroelectricity provides an additional control of this hybridized excitonic state. In 3L/3L $WSe_2$, the broken inversion symmetry and polarization-dependent stacking registry produce distinct h-IX branches with different energies and Stark responses, whereas such h-IX contrast is absent in the thinner 2L/2L structure despite clear ferroelectric domains in KPFM. Moreover, the hysteretic transfer of spectral weight between h-$IX^+$ and h-$IX^-$ directly tracks polarization switching under an external electric field. Together, these results show that layer hybridization governs the optical activity of the h-IX, while sliding ferroelectricity controls its layer polarization and optical response, establishing thickness and stacking polarization as complementary degrees of freedom for manipulating high-energy dipolar excitons in $WSe_2$ homostructures.

# Experimental section

### Device fabrication

WSe2 crystals were grown by chemical vapor transport in quartz ampoule directly from elements using bromine as a transport medium. Tungsten (99.999%, -100 mesh, China Rhenium Co, China) and selenium (99.9999%, 2-4 mm, Wuhan Xinrong New Materials Co, China) were placed in stoichiometric ratio corresponding to 50 g of WSe2 in quartz ampoule (50x250 mm) together with 2 at % excess of selenium and 1 g of SeBr4 (99.9%, STREM, USA) and melt sealed under high vacuum (<1x10-3 Pa using oil diffusion pump and LN2 cold trap). The ampoule was first heated in muffle furnace at 500 °C for 25 hours, on 600 °C for 50 hours and on 800 °C for 50 hours. The heating rate was 1 °C/min and free cooling. Subsequently the ampoule was placed in two zone furnaces for crystal growth. First the growth zone was heated to 1000 °C and source zone on 800 °C. After 2 days the thermal gradient is reversed, and source zone is heated to 1000 °C and growth zone on 900 °C for 14 days. Finally, the ampoule is cooled at room temperature with growth zone additional heated in 300 °C for 2 hours to remove volatile compounds. Finally, ampoule was opened in argon glovebox and crystals collected.

$WSe_2$ flakes were exfoliated using Gel-Film (DGL-45/17-X8), and the layer number was identified by optical contrast. Bilayer and trilayer $WSe_2$ flakes were transferred onto a $SiO_2$ (90 nm)/Si substrate at 70 °C. The devices presented in Supplementary Figure S1 were assembled using a dry polymer-stamping method.

**KPFM measurements**

KPFM images were acquired in an inert nitrogen environment using a Park NX-10 system operated in sideband mode. An HQ:NSC35/Pt tip (cantilever C, resonant frequency ≈ 150 kHz, spring constant 5.4 N $m^{-1}$) was used. For the sideband measurements, the resonant frequency of the tip was calibrated and the drive frequency was then set 2.5 kHz away from it. The tip was excited with 2 $V_{AC}$ at a 10 nm set-point for all measurements.

**TRPL measurements**

Time-resolved photoluminescence measurements were performed on the parallel-stacked 3L/3L $WSe_2$ device and reference 4L–6L 2H-$WSe_2$ regions at 8 K and room temperature using 470 nm pulsed excitation at 40 MHz and an average power of 0.3 μW. Emission was collected through a 100× objective with NA 0.75 and detected by two Si single-photon avalanche diodes coupled to a time-correlated single-photon-counting module, with 64 ps time bins. Spectral selection used either a fixed 540–580 nm bandpass filter or an angle-tuned 560–580 nm bandpass filter, whose nominal window shifted to approximately 542–562 nm at 30°. The instrument-response function used for comparison is shown in Fig. S3a.

**Optical characterization**

PL spectra were measured using a self-built reflection cryogenic microscope (attoDRY800 with an external objective – LUCPLFLN40X/0.6 NA). Measurements were performed with a linearly polarized 532 nm CW laser. The PL spectra were measured using a Shamrock 303i Spectrometer coupled with an iDus401 CCD camera (sensor cooled via thermoelectric cooling to 205 K). Keithley's 2450 SourceMeter Unit was used to set the device gate voltage and monitor leakage current through the hBN barrier.

## Acknowledgements

I.K. acknowledges support from the Azrieli Foundation through the Azrieli Graduate Studies Fellowship. M.B.S. acknowledges funding by the European Research Council under the European Union's Horizon 2024 research and innovation program ('SlideTronics', consolidator grant agreement no. 101126257) and the Israel Science Foundation under grant no. 319/22. We further acknowledge the Centre for Nanoscience and Nanotechnology of Tel Aviv University. J.-Y. You is supported by the National Natural Science Foundation of China (Grant No. 12604335Z.S. was supported by project LUAUS25268 from Ministr y of Education Youth and Sports (MEYS) and by funding from the European Union's Horizon Europe – European Innovation Council programme under grant agreement 101 223116. K.W. and T.T. acknowledge support from the JSPS KAKENHI (grant numbers 21H05233 and 23H02052) and World Premier International Research Center Initiative (WPI), MEXT, Japan.

# Hybridization-brightened interlayer excitons in multilayer $WSe_2$ controlled by sliding ferroelectricity

Supplementary Information

**Contents**

**List of figures**

**List of tables**

## S1. Device characterization

In the main text we present data from three selected devices. Device 1 is a parallel-stacked (0° twist angle) 3L + 3L $WSe_2$ homostructure, encapsulated by 20.44 nm-thick h-BN and ~10 nm-thick graphene, on a 90 nm-thick $SiO_2$/Si substrate. In device 2, we have used 17 nm thick and 13 nm thick h-BN encapsulating 3L/3L $WSe_2$. In device 3, 3,4,5,6 layer 2H $WSe_2$ is on 10 nm-thick graphene and top gate is on top of 28 nm thick h-BN. Optical image and layer thicknesses is presented in Supplementary figure 1.

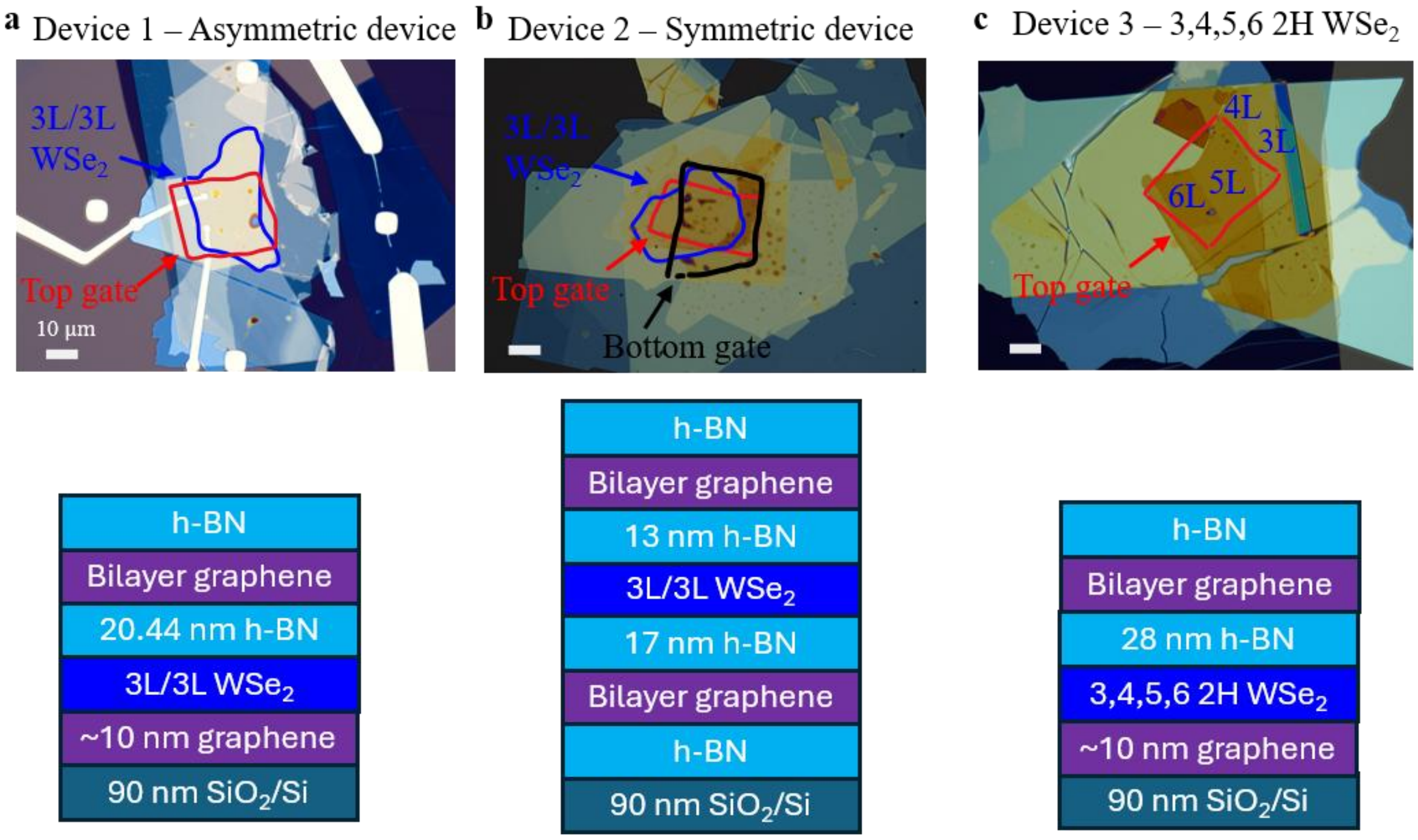

**Figure S1. Optical device images. a,b,c**, Optical microscope images of devices 1,2,3 used in the paper. Scale bars indicate 10 µm.

## S2. Linear fitting for Stark effects

To quantify the electric-field response of the exciton, we extracted the peak positions from the field-dependent maps in Fig. S2a (black squares) and fitted them with a linear function in the negative-field regime (red lines). The extracted slopes and their standard errors are summarized in Supplementary Fig. S2b. The 3L sample shows no discernible field dependence and is therefore excluded from the fitting. The slope, which reflects the out-of-plane dipole moment of the exciton, is largest for 4L (0.165 e·nm) and decreases for 5L (0.0848 e·nm) and 6L (0.107 e·nm). The small standard errors (~0.001 e·nm) confirm the reliability of the linear fits.

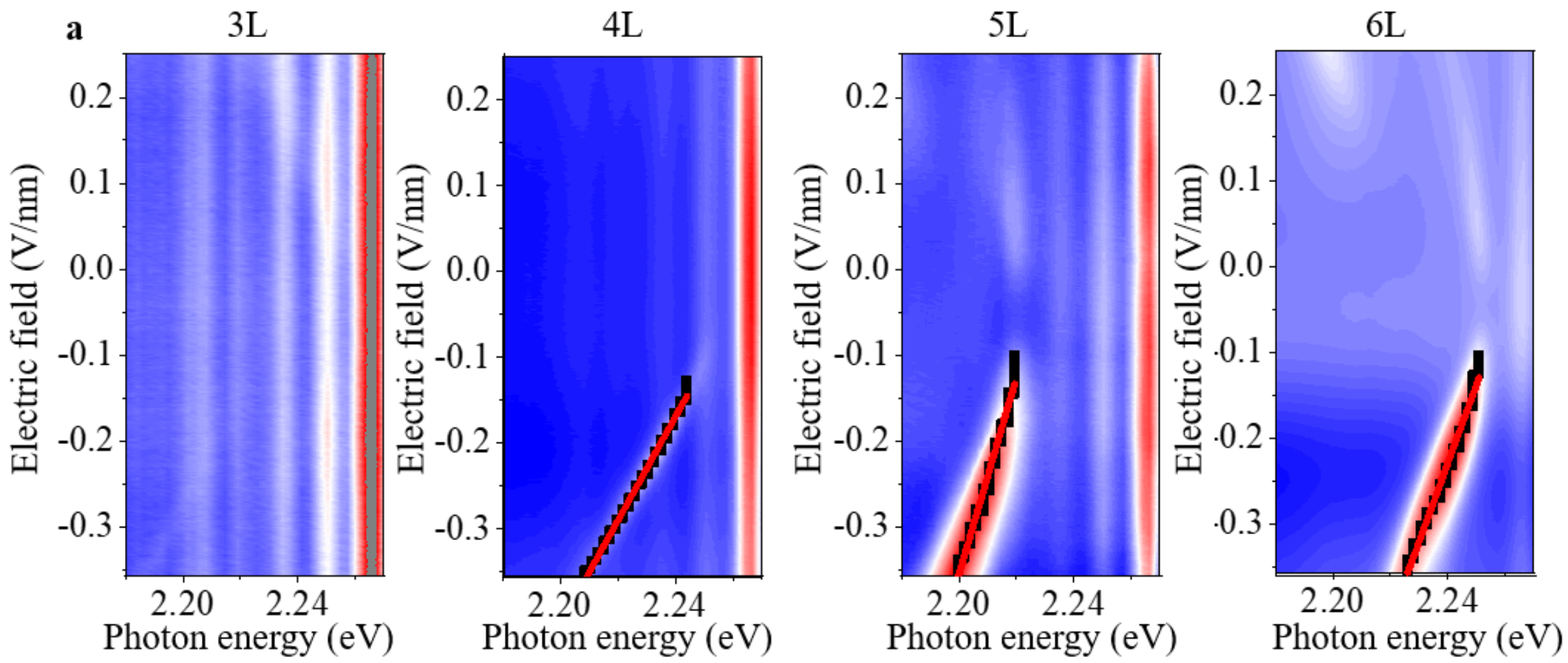

b

| | 4L | 5L | 6L |
|---|---|---|---|
| Slope (e·nm) | 0.165 | 0.0848 | 0.107 |
| Standard error | 0.00126 | 0.00114 | 0.00126 |

**Figure S2. Electric-field dependence of the exciton peak and linear fitting of the Stark shift. a**, Stark shifts for 3L, 4L, 5L, and 6L samples. Black squares mark the extracted peak positions, and red lines are the linear fits to the field-dependent peak shift in the negative-field regime. No measurable shift is resolved for 3L, so no fit is shown. **b**, Slope and standard error of the linear fits for 4L, 5L, and 6L.

## S3. Time resolved photoluminescence (TRPL) measurements

TRPL measurements were performed as described in Methods. At 8 K, the 0° and 10° traces from position PD1 in a downward-polarized domain of the 3L/3L device show prompt profiles close to the instrument-response function (Fig. S3a). The 20° and 30° signals are weaker relative to the background but retain a modest delayed contribution after baseline subtraction and peak normalization (inset). Exploratory biexponential-plus-offset fitting of a separate 30° trace, with the fast term fixed at 0.6 ns, yields an apparent slow timescale of order 5 ns. This value depends on the fitting choices. Neither decay component can be assigned unambiguously to the h-IX.

With fixed 540–580 nm detection, the 4L region exhibits a more pronounced delayed contribution extending over several nanoseconds than the 5L, 6L, and 3L/3L regions (Fig. S3b). Each trace is normalized to its reference baseline, so the comparison does not establish absolute emission yields. If the delayed emission involves states participating in h-IX formation or relaxation, its reduced contribution in thicker regions could reflect changes in their population or feeding. However, differences in spectral backgrounds and recombination pathways remain possible.

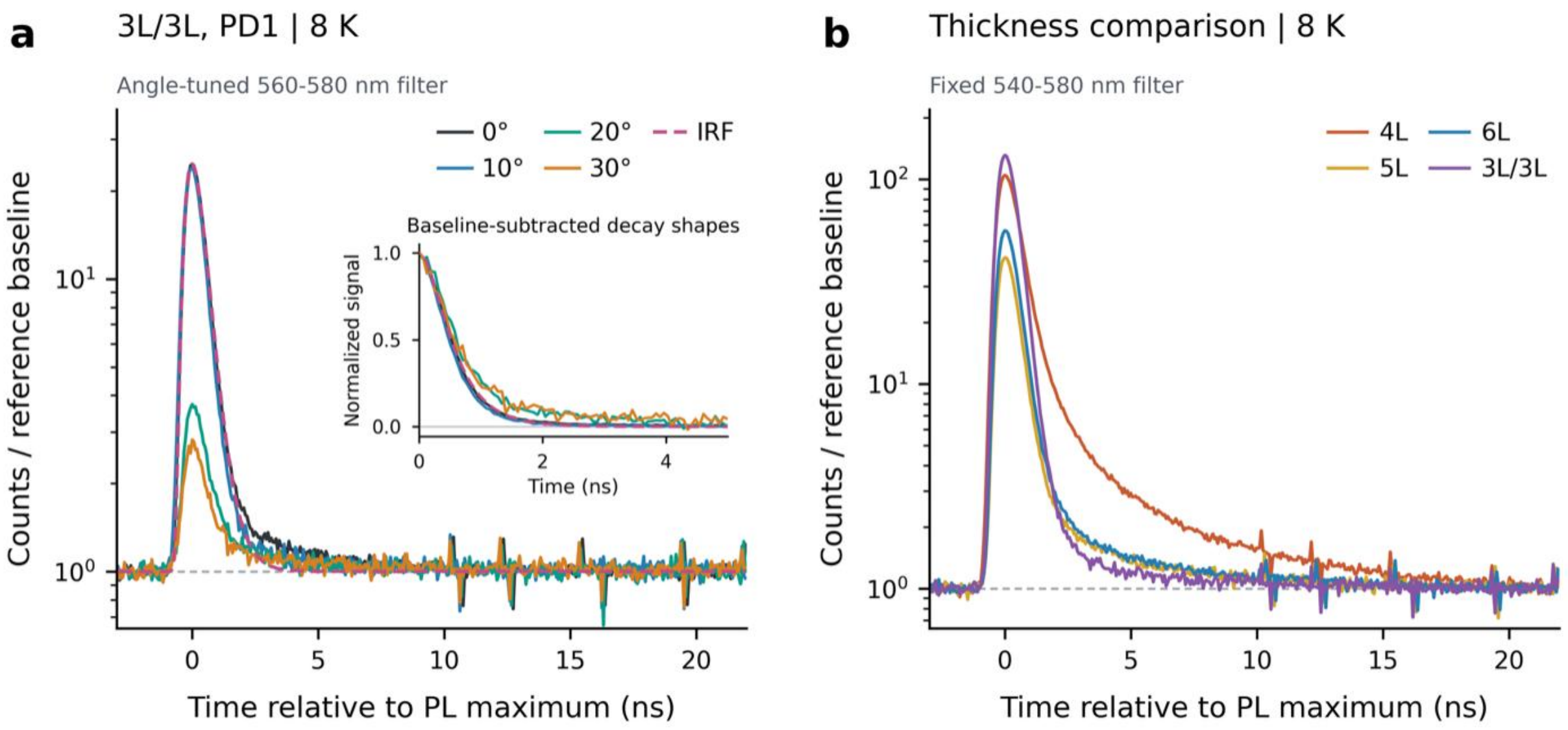

**Figure S3**. **Spectrally filtered time-resolved photoluminescence of $WSe_2$ at 8 K**. **a,** Traces from position PD1, a measurement point in a downward-polarized (P↓) domain of the 3L/3L device, recorded using the angle-tuned 560–580 nm filter at 0°–30°. The instrument-response function (IRF) is scaled to the peak and reference baseline of the 0° trace. Inset, the same traces after baseline subtraction and normalization to the remaining peak. **b,** Traces from 4L, 5L, and 6L 2H-$WSe_2$ and the 3L/3L structure using fixed 540–580 nm detection. Each main-panel trace is normalized to its reference baseline. Dashed lines mark unity. Time is referenced to the common PL maximum within each dataset.

## S4. GW-BSE calculations and exciton analysis

### S4.1 Electronic-structure and GW calculations

The 4L and 6L 2H-$WSe_2$ structures were fully relaxed within density-functional theory using Quantum ESPRESSO[1,2] and the Perdew–Burke–Ernzerhof (PBE) exchange-correlation functional[3]. Interlayer van der Waals interactions were included during structural optimization using the DFT-D3 dispersion correction[4]. The optimized in-plane lattice constants are approximately 3.32 Å for both thicknesses. Spin-orbit coupling was included using fully relativistic norm-conserving PBE pseudopotentials obtained from the Quantum ESPRESS O pseudopotential repository[5]. A plane-wave kinetic-energy cutoff of 60 Ry was used. Approximately 15 Å of vacuum was included normal to the layers, giving total supercell heights $L_z$ = 40.97 Å for 4L and 56.03 Å for 6L. The coarse wavefunction mesh used for G W-BSE was 18×18×1.

Quasiparticle energies were calculated within the GW approximation using BerkeleyGW[6–8]. A slab Coulomb truncation was applied along the nonperiodic direction to remove interactions between periodically repeated images[9]. The dielectric matrix was evaluated with a 20 Ry cutoff. The dielectric-function and self-energy summations included 400 bands for 4L and 600 bands for 6L, respectively. The resulting direct K-K quasiparticle gaps are 2.154 eV (4L) and 2.130 eV (6L), while the indirect K-Q gaps are 1.752 eV and 1.698 eV, respectively.

### S4.2 Bethe-Salpeter equation and optical spectra

Neutral excitations were obtained by solving the zero-center-of-mass-momentum Bethe-

Salpeter equation (BSE) in the Tamm-Dancoff approximation, including the statically screened direct electron-hole attraction and the bare exchange interaction[8,10,11]. The active BSE subspaces contain 8 valence plus 8 conduction bands for 4L and 12 valence plus 12 conduction bands for 6L. The electron-hole kernel was constructed on the 18×18×1 coarse mesh and interpolated to a 36×36×1 fine mesh for the optical calculation. Spinor wavefunctions were used throughout. The velocity matrix elements were evaluated using a shifted fine-grid wavefunction with $q_{\text{shift}}$ = (0.001, 0.002, 0) in crystal coordinates, defining the in-plane finite-q direction used in the $q \rightarrow 0$ optical limit. Fine-grid and shifted fine-grid spinor wavefunctions were explicitly evaluated on all 36×36×1 points without fine-grid symmetry reduction. The plotted spectra use a Voigt kernel with Gaussian standard deviation $\sigma$ = 0.020 eV and Lorentzian half-width $\gamma$ = 0.020 eV (total FWHM ≈ 72.0 meV). The same interpolation, broadening, and solver settings were used for 4L and 6L.

The sheet optical response was reconstructed from the 10,000 computed exciton states in each system using the native BerkeleyGW normalization:

$$L_z\varepsilon_2(E)=L_zC\sum_S F_S[V(E-\Omega_S)-V(E+\Omega_S)], \qquad F_S=|d_S|^2.$$

Here, $V$ is the unit-area Voigt kernel and $C$ is the native BerkeleyGW volume and spinor normalization prefactor. Multiplication by $L_z$ gives the sheet optical response. In the native dipole convention, $L_zC \approx 0.01287$ for both systems, with $E$ in eV and $L_z$ in Å. $F_S$ is the squared transition dipole reported by the optical calculation.

The peak-centred ±60 meV windows are 2.2771–2.3971 eV for 4L and 2.2878–2.4078 eV for 6L, containing native states 671-1214 and 1519-2791, respectively. The discrete window sum and the finite-window integral of the broadened sheet response are defined as:

$$F_D = \sum_{S\in D} F_S, \qquad S_{2D} = \int_{E_p-\Delta E}^{E_p+\Delta E} L_z\varepsilon_2(E)\,dE, \quad \Delta E = 0.060 \text{ eV}.$$

The 6L/4L ratios are 1.46 for $F_D$ and 1.47 for $S_{2D}$; the broadened integral includes tails from states outside the discrete window.

**Table S1. State counts and summed squared transition dipoles in the analysis windows.** The interlayer-dominated subset is defined by $W_{\text{inter}} > W_{\text{intra}}$. All strengths use the same native dipole convention.

| Quantity | 4L | 6L |
|---|---|---|
| All states in window | 544 | 1,273 |
| Interlayer-dominated states | 445 | 1,097 |
| Total window sum $F_D$ | 5232.59 | 7649.51 |
| Interlayer-dominated sum | 523.77 | 2408.98 |
| Interlayer-dominated optical share | 10.01% | 31.49% |

S7.3 Momentum-space distribution and GW-band projections

The zero-momentum exciton is expanded in spinor conduction and valence bands as

$$|S\rangle = \sum_{vc\mathbf{k}} A^S_{vc\mathbf{k}} |c\mathbf{k}; v\mathbf{k}\rangle.$$

The band indices v and c label spinor states including spin–orbit coupling. For the analysis window D, the transition-pair distribution is averaged with squared-dipole weights:

$$W_D(k,v,c)=\frac{\sum_{S\in D}F_S|A^S_{vck}|^2}{\sum_{S\in D}F_S}, \qquad \sum_{kvc}W_D(k,v,c)=1.$$

$$W_D(k)=\sum_{vc}W_D(k,v,c), \qquad W^e_{ck}=\sum_v W_D(k,v,c), \qquad W^h_{vk}=\sum_c W_D(k,v,c).$$

Figure 2b displays the resulting momentum distribution and the marginal weights on Wannier-interpolated GW quasiparticle bands[12]. The energy zero is the valence-band maximum at K. The Q minima lie approximately halfway from Γ to K. The optical configurations in both the K/K′ and Q regions are vertical transitions, with the electron and hole at the same crystal momentum.

To resolve the optical contribution of each valley, the native BSE mesh was divided into K, Q and remaining regions. All symmetry-related pockets within 0.25|ΓK| of each centre were included, with overlapping points assigned to the nearest pocket. Parabolic interpolation of the native conduction energies gives |ΓQ|/|ΓK| = 0.509 for 4L and 0.507 for 6L. The optical amplitude of region α is obtained by restricting the native optical contraction to **k** points in the set $\Omega_\alpha$:

$$d^S_\alpha = \sum_{\mathbf{k}\in\Omega_\alpha}\sum_{v,c}A^S_{vc\mathbf{k}}\,g^*_{vc\mathbf{k}}, \qquad F_S = |\sum_\alpha d^S_\alpha|^2, \qquad \alpha \in \{\mathrm{K}, \mathrm{Q}, \mathrm{other}\}.$$

Here v and c label valence and conduction bands, respectively, S labels the exciton state, and g is the native complex optical vector. The set $\Omega_\alpha$ contains the **k** points assigned to region α, with "other" denoting the remaining region. The optical self term of region α is the window sum of $|d_\alpha^S|^2$, and the K–Q cross term is the window sum of $2\mathrm{Re}[d_K^S(d_Q^S)^*]$. The remaining-region self term and its interference with K and Q complete the optical sum. Joint valley–layer contributions use the layer matrices defined below, with the same restriction on **k**. Table S2 distinguishes the optical-strength-weighted pair probabilities from the optical self and cross terms.

**Table S2. Valley and layer contributions in the analysis windows.** Pair probabilities are averaged with $F_S$ weights. Optical self and cross terms are divided by the total window dipole strength; K and Q include all symmetry-related pockets. The Q-intralayer optical self term is a component of the Q-sector decomposition and is not added again to the valley terms above it.

| Quantity (%) | 4L | 6L |
|---|---|---|
| K-region pair probability | 19.33 | 41.74 |
| Q-region pair probability | 80.01 | 57.85 |
| K optical self term | 3.82 | 8.16 |
| Q optical self term | 76.49 | 71.11 |

| Quantity (%) | 4L | 6L |
|---|---|---|
| K–Q coherent cross term | 12.22 | 14.09 |
| Q-intralayer optical self term | 68.83 | 63.40 |
| K-interlayer pair probability | 12.99 | 30.41 |
| Q-intralayer pair probability | 63.05 | 43.59 |

**S4.4 Global electron–hole layer probabilities**

For each normalized exciton, the real-space spinor amplitude is

$$\Psi^{S}_{\sigma_e\sigma_h}(r_e,r_h)=\sum_{vck}A^{S}_{vck}\psi_{ck,\sigma_e}(r_e)\psi^{*}_{vk,\sigma_h}(r_h).$$

The global joint layer probability integrates over both particle coordinates and sums all four spinor-component pairs. Writing $A_S(k)$ as a matrix with conduction rows and valence columns, it is evaluated from the band layer matrices $C_i(k)$ and $V_j(k)$:

$$P^{S}_{ij}=\sum_k \mathrm{Tr}[A^{\dagger}_{S}C_iA_SV_j].$$

Here $C_{i,cc'} = \langle ck|\chi_i(z)|c'k\rangle$ and $V_{j,vv'} = \langle vk|\chi_j(z)|v'k\rangle$, with $\chi_i$ the indicator of layer i; k arguments are suppressed in the trace. The layer matrices sum both spinor components, and all off-diagonal band elements and complex phases are retained. Matrix elements are obtained by analytical integration of plane-wave products over each z interval.

Layers are numbered from top to bottom, with $L1$ at the largest z. Internal boundaries are midpoints between neighbouring W planes; outer boundaries extend half the neighbouring W-plane spacing beyond the outer W plane. The remaining part of the simulation cell forms a residual domain. The probabilities are normalized over the full cell, with no renormalization after excluding that domain:

$$W_{intra}=\sum_i P^{S}_{ii}, \qquad W_{inter}=\sum_{i\neq j}P^{S}_{ij}, \qquad W_{res}=1-W_{intra}-W_{inter}.$$

The two material-layer sums run over the $WSe_2$ layers. $W_{res}$ includes configurations with at least one particle in the residual domain.

The representatives in Fig. S4 are the strongest interlayer-dominated states in their windows: native state 828 at 2.31424 eV for 4L and state 2183 at 2.34213 eV for 6L. Their intralayer/interlayer probabilities are 14.89%/85.08% and 37.56%/62.41%, respectively. The optical-strength-weighted global interlayer probabilities of the complete windows are 29.76% for 4L and 44.62% for 6L.

### S4.5 Coherent optical decomposition

Let g(k) denote the native optical vector, with $d_S = \langle g, A_S \rangle$. Applying the same layer matrices to the exciton amplitude defines the contribution of each electron–hole layer pair:

$$d_{ij}^{S} = \sum_k \mathrm{Tr}[g^\dagger C_i A_S V_j].$$

The diagonal material-layer amplitudes are summed into dintra, the off-diagonal material-layer amplitudes into dinter, and terms involving the residual domain into dres. Their coherent sum recovers the total amplitude:

$$d = d_{intra} + d_{inter} + d_{res},$$

$$F = |d_{intra}|^2 + |d_{inter}|^2 + 2\mathrm{Re}(d_{intra} d_{inter}^{*}) + R,$$

$$R = |d_{res}|^2 + 2\mathrm{Re}[(d_{intra} + d_{inter}) d_{res}^{*}].$$

Optical amplitudes are added coherently within each exciton, followed by a sum of the self and cross terms over states in the analysis window. The residual R contains its self term and its interference with material-layer amplitudes and can have either sign.

**Table S3. Coherent optical decomposition in common native dipole-strength units.** The full-window increase is 2416.92. Its intralayer, interlayer, cross and residual contributions are 90.11%, 0.33%, 9.68% and −0.12%, respectively. Cross and residual terms retain their signs.

| Selection | Intra self | Inter self | Cross | Residual | Total |
|---|---|---|---|---|---|
| 4L window | 4745.21 | 12.24 | 439.53 | 35.61 | 5232.59 |
| 6L window | 6923.10 | 20.24 | 673.37 | 32.80 | 7649.51 |
| 4L state 828 | 29.85 | 0.07 | 2.82 | −0.01 | 32.73 |
| 6L state 2183 | 351.31 | 1.71 | 49.05 | −0.29 | 401.79 |

For the representative states, the total strength increases by a factor of 12.27, with 87.10% of the increase in the intralayer self term and 12.53% in the cross term. For the interlayer-dominated subsets, the corresponding fractions are 90.68% and 8.94%. Within these subsets, the Q-region intralayer self term carries 55.44% of the total dipole strength in 4L and 61.63% in 6L, making it the largest optical self contribution in both systems. The mixed states carry optical strength through their intralayer amplitudes as well as the coherent cross term.

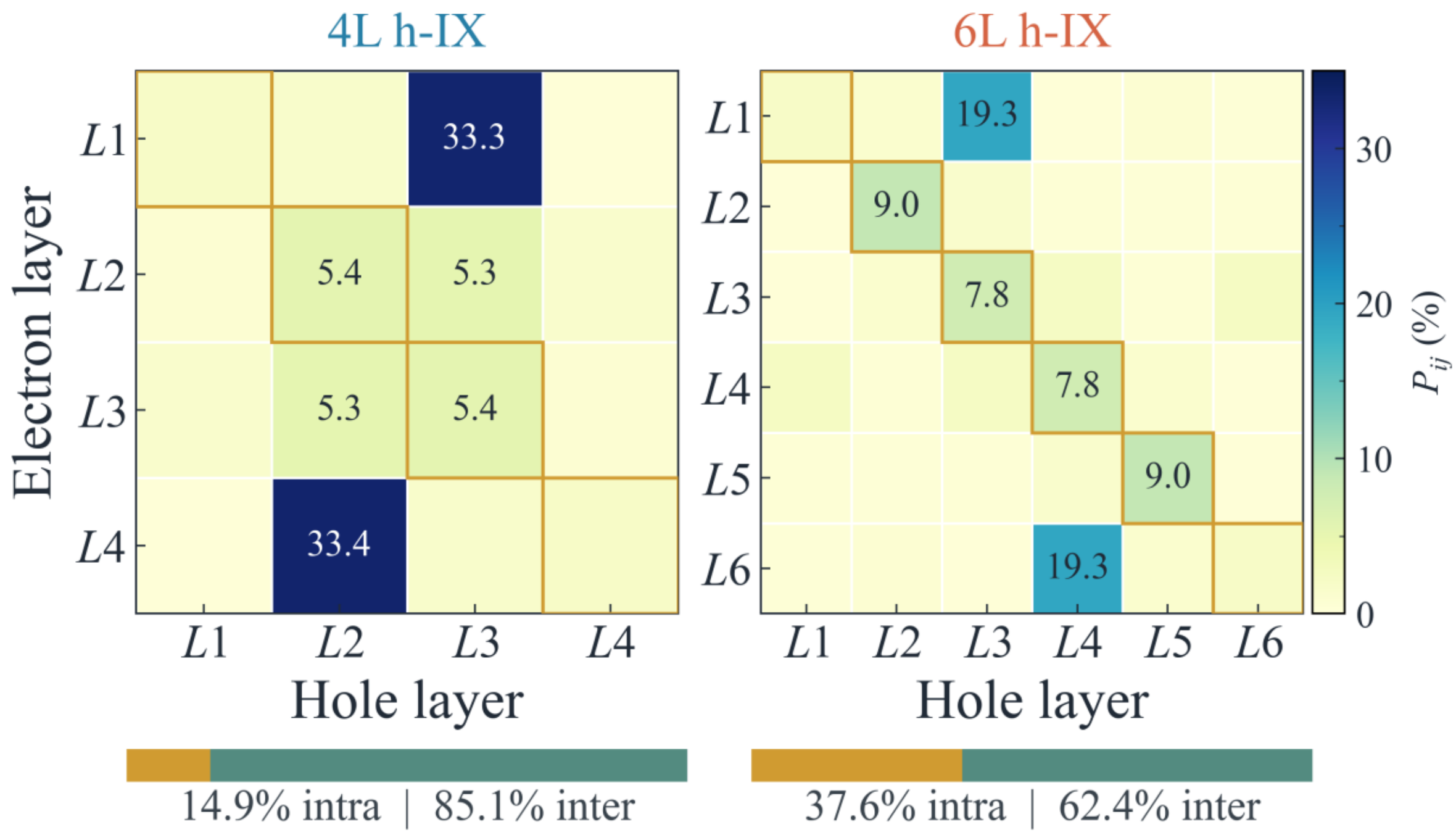

**Figure S4. Global joint electron–hole layer probabilities of representative 4L state 828 and 6L state 2183, integrated over both particle coordinates and all spinor components.** Rows and columns denote electron and hole layers; outlined diagonal cells identify intralayer configurations. Bars summarize intralayer and interlayer probabilities. These states are the strongest interlayer-dominated state in each analysis window; their optical decomposition is shown in Fig. 2f.

### S4.6 Frozen-basis model of the Stark response

The field response was estimated in the zero-field BSE basis, retaining 2,029 states for 4L (native indices 185–2213) and 4,534 states for 6L (311–4844), spanning approximately 2.10–2.55 eV. The quasiparticle orbitals, screening, interaction kernel and optical vertex are held fixed, while the field mixes states within this basis:

$$H_{ST}(F_{int})=\Omega_S^0\delta_{ST}+eF_{int}D_{ST},$$

$$D_{ST}=\langle S|z_e-z_h|T\rangle=\sum_k \mathrm{Tr}[A_S^{\dagger}(Z^c A_T-A_T Z^v)].$$

$Z^c$ and $Z^v$ are the spinor-band position matrices, $\langle a|z|b\rangle$ in Å, with the periodic coordinate cut in the vacuum. With energies in eV, Fint in V nm$^{-1}$ and D in Å, the field term is 0.1 Fint D. The sign convention uses $D = z_e - z_h$; effective lengths use the magnitude of the fitted slope.

At each field, model eigenvalues are sorted from lowest to highest, with $E_{(n)}$ denoting the nth eigenvalue. Selected energy-ordered levels are fitted by ordinary least squares over $-0.15 \leq$ Fint $\leq -0.06$ V nm$^{-1}$, with a 0.0025 V nm$^{-1}$ step (37 points):

$$E_{(n)}(F_{int})=a_n+b_nF_{int}, \qquad d_{eff,n}[\text{Å}]=10|b_n[\text{eV}/(\text{Vnm}^{-1})]|.$$

The selected zero-field references are native states 828–829 for 4L and 2181–2188 for 6L, including the representative h-IX states and their nearby levels. All retained basis states enter the finite-field diagonalization. Each curve uses the same energy-order index n at every field, while its excitonic composition can change. Each point in Fig. 2d reports the fitted length of one model level (Table S4).

**Table S4. Selected energy-ordered model levels and their effective Stark lengths over -0.15 to -0.06 V nm$^{-1}$.** Native indices and energies identify the zero-field BSE states. The maximum fit residual measures the departure from a linear energy shift.

| System | Model level n | Zero-field BSE state | Zero-field energy (eV) | $d_{eff}$(Å) | Maximum fit residual (meV) |
|---|---|---|---|---|---|
| 4L | 644 | 828 | 2.31424 | 3.45 | 4.38 |
| 4L | 645 | 829 | 2.31427 | 3.45 | 4.20 |
| 6L | 1871 | 2181 | 2.34203 | 2.76 | 1.73 |
| 6L | 1872 | 2182 | 2.34203 | 2.76 | 1.86 |
| 6L | 1873 | 2183 | 2.34213 | 2.74 | 1.85 |
| 6L | 1874 | 2184 | 2.34216 | 2.73 | 1.85 |
| 6L | 1875 | 2185 | 2.34222 | 2.73 | 1.83 |
| 6L | 1876 | 2186 | 2.34223 | 2.73 | 1.90 |
| 6L | 1877 | 2187 | 2.34228 | 2.72 | 1.92 |
| 6L | 1878 | 2188 | 2.34229 | 2.72 | 1.73 |

The internal field is $F_{int} = (\varepsilon_{hBN}/\varepsilon_{WSe2})\, E_z = (3/7)\, E_z$. The experimental device-field slopes of 0.165 and 0.107 e·nm therefore correspond to 3.85 and 2.50 Å for 4L and 6L, on the same internal-field convention as the model. The common model interval, −0.15 to −0.06 V nm$^{-1}$ in internal field, therefore corresponds to −0.35 to −0.14 V nm$^{-1}$ in device field.

The fitted lengths characterize the selected model levels over this common field interval; the underlying dispersions are nonlinear (Fig. S5).

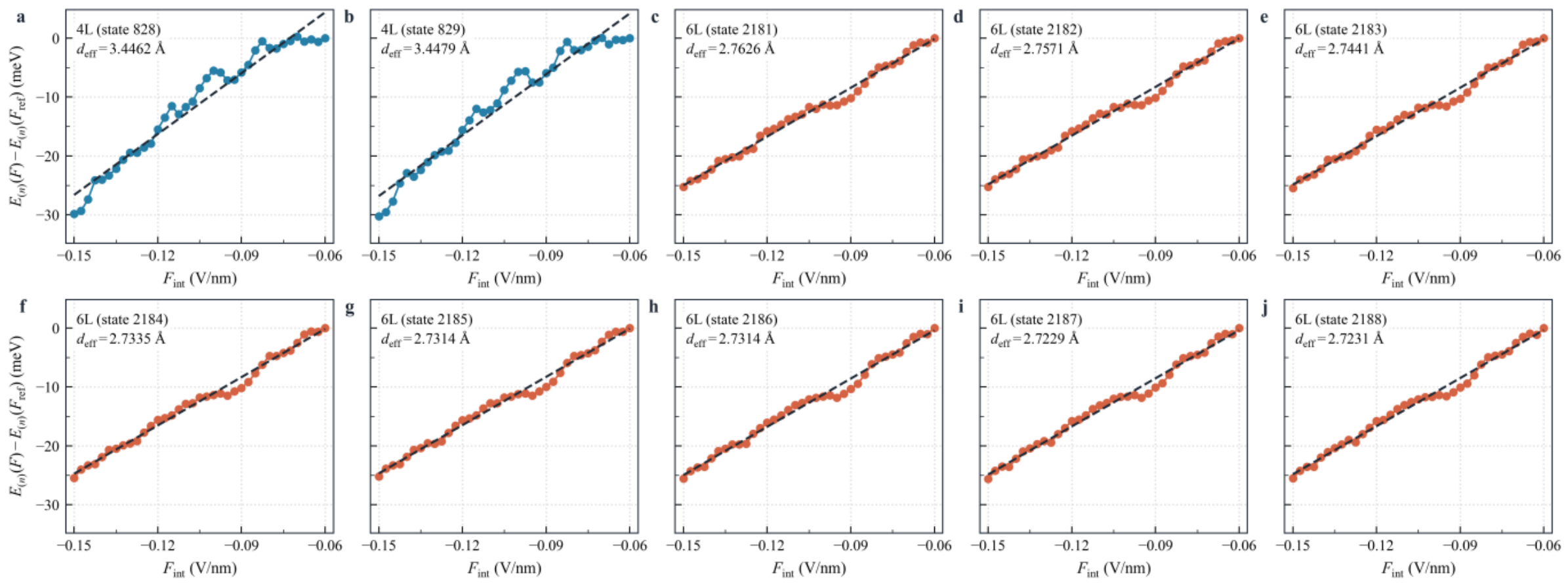

**Figure S5. Model energy shifts and linear fits for the selected energy-ordered model levels n listed in Table S4.** The common internal-field interval is −0.15 to −0.06 V nm$^{-1}$. Energy shifts are referenced to each model level's ene rgy at −0.06 V nm$^{-1}$. Each fitted curve provides one model point in Fig. 2d. Panel labels iden tify the corresponding zero-field BSE states.

## S5. Comparison of PL peak resolution

In 3L/3L $WSe_2$, we observed three peaks: the A exciton, the h-IX, and the UPL. To compare the resolution of these peaks for the $P_{up}$ and $P_{down}$ domains, we inserted a graphene spacer between the two 3L $WSe_2$ flakes. The $P_{up}$ and $P_{down}$ domains are distinguished in the KPFM image (Fig. S6a). The integrated h-IX map (2.222 eV – 2.242 eV) shows that $P_{up}$ and $P_{down}$ remain distinguishable down to the 600 nm circles and the 300 nm × 1500 nm bar-shaped cavities. This confirms that the peak shift of the interlayer exciton is well suited to resolving small cavities, offering higher spatial contrast than the A exciton and UPL channels.

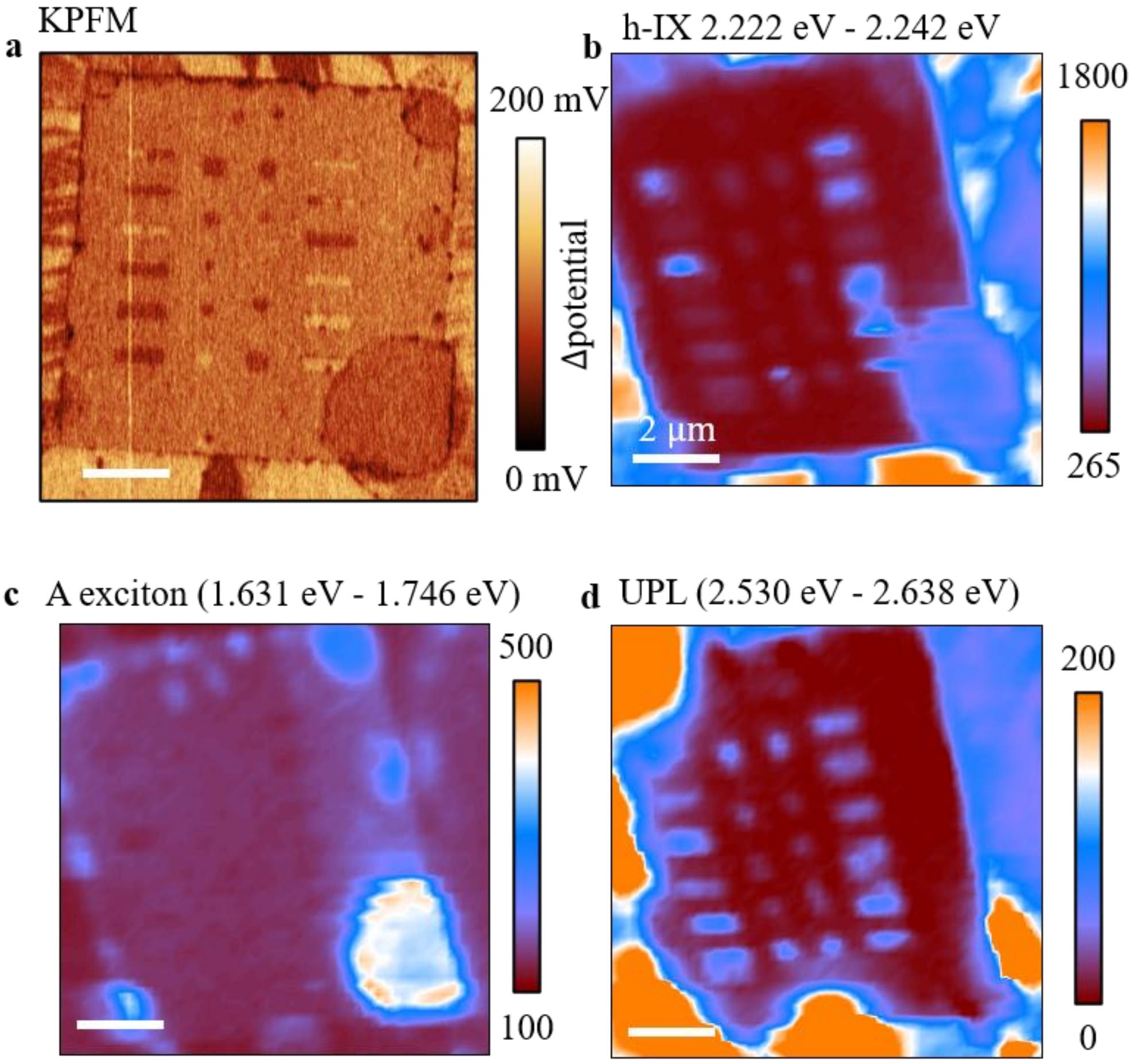

**Figure S6. Comparison of PL peak resolution for distinguishing sliding ferroelectric domains. a,** KPFM image of the 3L/3L $WSe_2$ structure with an inserted graphene spacer, showing the $P_{up}$ and $P_{down}$ domains via their surface potential contrast (Δpotential, 0-200 mV range). Sub-micron bar- and circle-shaped cavities are patterned to confine the sliding ferroelectric domains. **b,** Spatial map of the integrated h-IX emission (2.222 eV – 2.242 eV), which resolves $P_{up}$ and $P_{down}$ domains down to the 600 nm circles and 300 nm × 1500 nm bars. **c**, Spatial map of the integrated A exciton emission (1.631 eV – 1.746 eV). **d**, Spatial map of the integrated UPL emission (2.530 eV – 2.638 eV). All scale bars, 2 μm.

## S6. Characterization of 2L/2L $WSe_2$

For comparison with the 3L/3L $WSe_2$ PL spectrum, we fabricated a 2L/2L $WSe_2$ sample with a nearly 0° 3R interface. The KPFM image confirms that this contrast corresponds to polarization domains (Fig. S7a). The line profile gives a surface potential difference of ~100 mV (Fig. S7b). Point spectra extracted from the $P_{up}$ and $P_{down}$ domains (Fig. S7c) show that the $KK_A$ peak differs by ~10% in intensity between the two domains, whereas the UPL signal differs by nearly a factor of four. This demonstrates that UPL provides a larger intensity contrast between polarization domains than the $KK_A$ emission

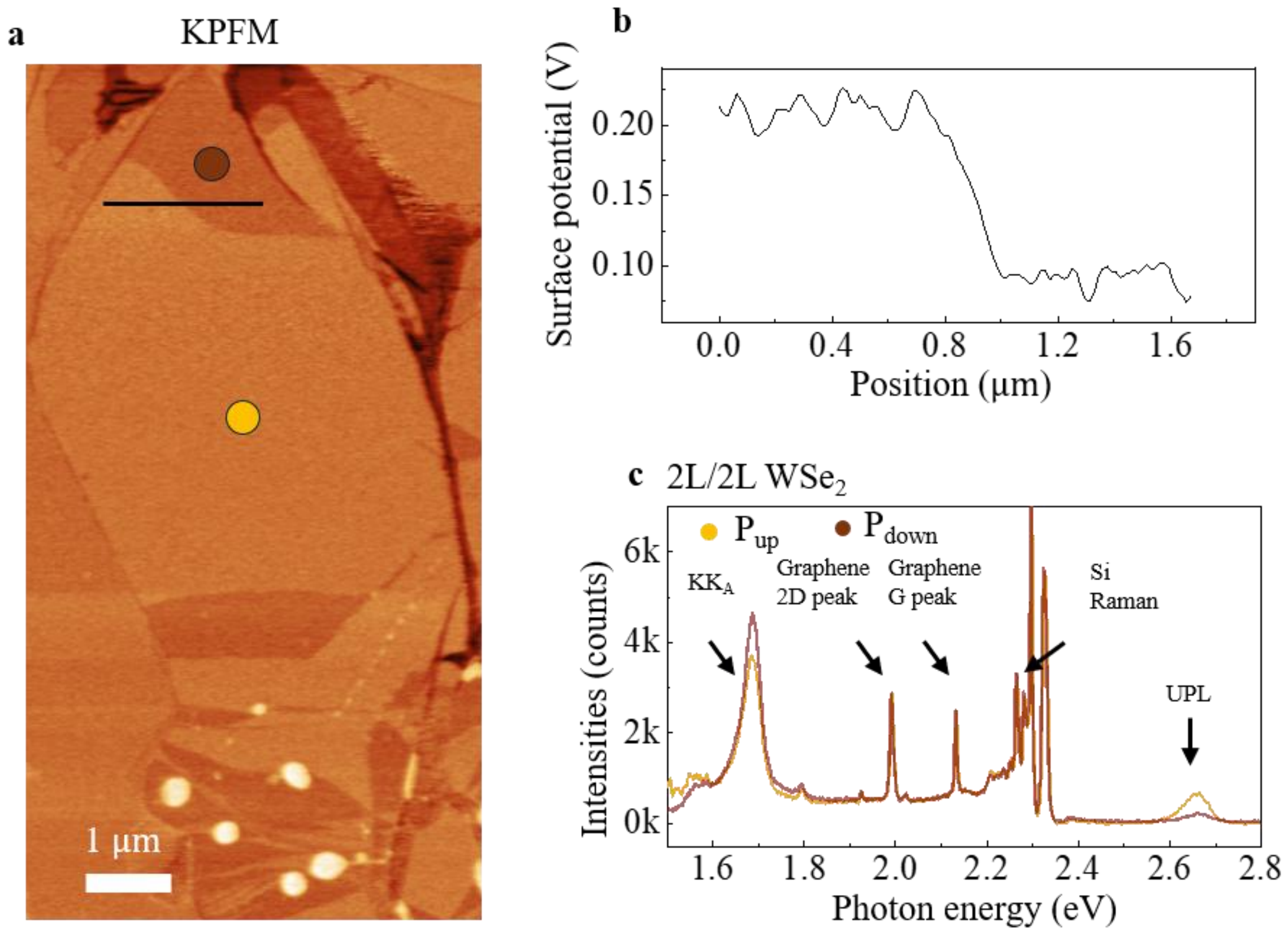

**Figure S7. Characterization of 2L/2L $WSe_2$. a,** KPFM image of 2L/2L $WSe_2$. **b**, KPFM line profile from the black guideline in **a**, **c**, PL spectrum extracted from yellow and brown area marked in **a**.

## S7. h-IX in symmetric device structure

To confirm that the h-IX emission originates from the sliding ferroelectric polarization of the 3R-stacked $WSe_2$ rather than from an external field artifact, we characterized a symmetric device in which the $WSe_2$ homostructure is under the top and bottom graphene gates with h-BN insulating layers (13 nm and 17 nm). This dual-gate geometry allows the vertical displacement field and the carrier doping to be tuned independently.

KPFM imaging performed before the top-gate deposition resolves a network of ferroelectric domains with alternating surface potential (Fig. S8a), consistent with the $P_{up}$ and $P_{down}$ polarization states expected for marginally twisted 3R $WSe_2$. Unlike asymmetric device, PL is indistinguishable between $P_{up}$ and $P_{down}$ domains. Spatially resolved PL mapping over the h-IX emission window (550–580 nm, ≈2.138–2.254 eV) are reproduced under a vertical electric field (Fig. S8b)

Point spectra from the $P_{up}$ and $P_{down}$ regions (Fig. S8c) show that the h-IX peak near 2.23 eV emerges when the applied field is parallel to the local polarization. Applying the field from bottom to top intensifies the $P_{up}$ domain, whose dipole is likewise oriented from bottom to top, while suppressing the $P_{down}$ domain.

The field-dependent PL colormaps (Fig. S8d) reveal a linear Stark dispersion of the h-IX peak, with the $P_{up}$ and $P_{down}$ domains dispersing in opposite directions. Sweeping the field traces out a hysteretic switching loop, confirming the reversible ferroelectric switching of the polarization.

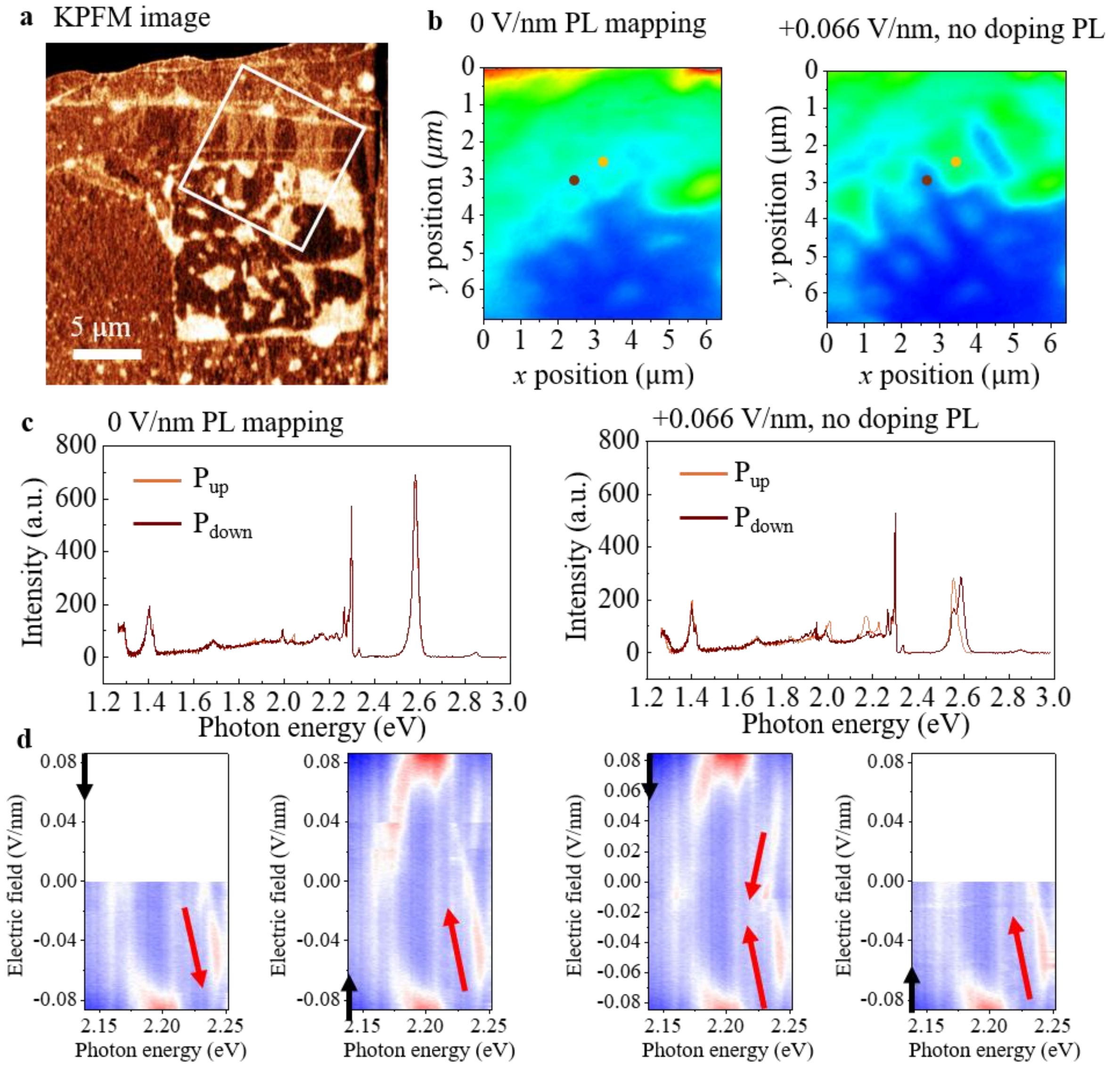

**Figure S8. h-IX mapping and polarization switching in a symmetric device.** Device structure: h-BN / bilayer graphene (BLG) / 13 nm h-BN / 3L $WSe_2$ / 3L $WSe_2$ / 17 nm h-BN / few-layer graphene / h-BN. **a,** KPFM map of the symmetric device acquired before deposition of the top graphene gate, revealing the ferroelectric domain pattern (orange ($P_{up}$) /brown ($P_{down}$) colors) of the 3R-stacked 3L+3L $WSe_2$. **b,** PL intensity mapping filtered over 545–560 nm (≈2.21–2.28 eV, the h-IX emission band) at zero gate voltage (left) and at a total field of 0.066 V/nm satisfying the charge-neutral condition $E_{bg} = -E_{tg}$ (right). **c,** Point spectra extracted from the $P_{up}$ (orange) and $P_{down}$ (brown) domains marked in **b**. **d,** Electric-field-dependent PL colormaps (2.15–2.25 eV) showing the Stark dispersion of the h-IX peak; red arrows indicate

the shift direction under sweeping field, and black arrows mark the sweeping directions.